\documentclass[twocolumn]{elsart3p}
\usepackage{graphicx} 
\usepackage{dcolumn}  
\usepackage{longtable}
\usepackage{bm}       
\usepackage{subfigure}
\usepackage{subeqn}
\begin{document}
\begin{frontmatter}
\title{Separating Signals from Non-Interfering Backgrounds 
  using \\Probabilistic Event Weightings.}

\author{M. Williams},
\author{M. Bellis} and 
\author{C. A. Meyer} 
\address{Carnegie Mellon University, Pittsburgh PA, 15213}
\date{\today}

%
%
\begin{abstract} 
A common situation in experimental physics is to have a signal which can not
be separated from a non-interfering background through the use of any cut.
In this paper, we describe a procedure for determining, on an event-by-event 
basis, a quality factor ($Q$-factor) that a given event originated from 
the signal distribution. This $Q$-factor can then be used as an event weight
in subsequent analysis procedures, allowing one to more directly access the
true spectrum of the signal.
\end{abstract}

\end{frontmatter}
%

\section{\label{section:intro}Introduction}

In many physics analyses, one has non-interfering background events which 
cannot be cleanly separated from the desired signal. 
This is of particular concern when the problem being studied is 
multi-dimensional, {\em i.e.} there are kinematic correlations in the signal
which must be preserved.
One typical way to handle these situations is with the 
{\em side-band subtraction} method.
Distributions constructed using events outside the signal region are
subtracted from those using events inside the signal region to create
distributions in which the backgrounds have been removed.
While this method can be effective in some situations, implementing it can
become problematic if the kinematics of the background region are different
than those of the signal or if the problem is sufficiently multi-dimensional
such that binning the data is severely limited by statistical uncertainties.

In this paper, we describe a procedure for assigning each event in a data 
sample a quality factor ($Q$-factor) which gives the chance that it originates 
from 
the signal sample. This $Q$-factor can then be used as an event weight
in subsequent analysis procedures to gain access to the signal distribution.
For example, the $Q$-factors can be used to perform event-based unbinned 
maximum likelihood fits on the data to extract physical observables.
Eliminating the need to bin the data is highly desirable for the case of 
multi-dimensional problems. 
We note here that this procedure is not appropriate for situations where there 
are quantum mechanical interference effects between the signal and background.
\section{\label{section:method}Quality Factor Determination}

Consider a data set composed of $n$ total events, each of which is described 
by $m$ coordinates $\vec{\xi}$ (${m \geq 2}$).
The coordinates can be masses, angles, energies, {\em etc}. 
Furthermore, the data set consists of $n_s$ signal
and $n_b$ background events. Both the signal and background distributions 
are functions of the coordinates, $S(\vec{\xi})$ and $B(\vec{\xi})$ 
respectively. 
For this procedure, we need to know {\em a priori} the functional dependence
(possibly with unknown parameters) of
the signal and background distributions in terms of one of the 
coordinates.
We will refer to this coordinate as the reference coordinate and label it
$\xi_r$.

As an example, consider the case where the reference coordinate is a
mass. The functional dependence of the signal, in terms of $\xi_r$, might be 
given by a Gaussian or Breit-Wigner distribution. 
The background may be well represented
by a polynomial. In both cases, there could be unknown parameters
({\em e.g.} the width of the Gaussian); these are permitted when using this
procedure. No other {\em a priori} information is required concerning the 
dependence of $S(\vec{\xi})$ or $B(\vec{\xi})$ on any of the other 
coordinates.

The aim of this procedure is to assign each event a quality factor, or 
$Q$-factor, which gives the chance that it originates from the signal sample.
We first need to define a metric for the space spanned by $\vec{\xi}$
(excluding $\xi_r$). A reasonable choice is to use 
$\delta_{kl}/\mathcal{R}^2_k$ where $\mathcal{R}_k$ is the maximum possible 
difference between the coordinates $\xi_k$ of any two events in the sample. 
Using this metric, the distance between any two events, $d_{ij}$, 
is given as
\begin{equation}
  \label{eq:dist}
  d^2_{ij} = \sum\limits_{k \neq r} \left[ \frac{\xi^i_k - \xi^j_k}
    {\mathcal{R}_k} 
    \right]^2,
\end{equation}
where the sum is over all coordinates except $\xi_r$.

For each event, we compute the distance to all other events in the data
set, and retain the $n_c$ {\em nearest neighbor} events, including the events 
itself, 
according to (\ref{eq:dist}). The value of $n_c$, which varies depending on the
analysis, is discussed below. The $n_c$ events are then fit using the 
unbinned maximum likelihood method to obtain estimators for the parameters, 
$\vec{\alpha}$, in the probability distribution function
\begin{equation}
  F(\xi_r,\vec{\alpha}) = 
  \frac{F_s(\xi_r,\vec{\alpha}) + F_b(\xi_r,\vec{\alpha})}
  {\int\left[F_s(\xi_r,\vec{\alpha}) + F_b(\xi_r,\vec{\alpha})\right]d\xi_r},
\end{equation}
where $F_s$ and $F_b$ describe the functional dependence on the reference
coordinate, $\xi_r$, of the signal and background respectively.

The $Q$-factor for each event is then calculated as
\begin{equation}
  \label{eq:q-factor}
  Q_i = \frac{F_s(\xi^i_r,\hat{\alpha}_i)}
  {F_s(\xi^i_r,\hat{\alpha}_i) + F_b(\xi^i_r,\hat{\alpha}_i)},
\end{equation}
where $\xi_r^i$ is the value of the event's reference coordinate and 
$\hat{\alpha}_i$ are the estimators for the parameters obtained from the
event's fit.

If one wants to bin the data, the signal yield in a bin is obtained as
\begin{equation}
  \label{eq:sig-yield}
  \mathcal{Y} = \sum\limits_i^{n_{bin}} Q_i,
\end{equation}
where $n_{bin}$ is the number of events in the bin. For example, to
construct a histogram (of any dimension) of the signal, one would simply 
weight each event's contribution by its $Q$-factor. 
\section{\label{section:errors}Error Estimation}

It is also important to extract the uncertainties on the individual 
$Q$-factors so that we can obtain error estimates on measurable 
quantities.
The full covariance matrix obtained from each event's fit, $C_{\alpha}$, can
be used to calculate the uncertainty in $Q$ as
\begin{equation}
  \label{eq:sigma-q}
  \sigma^2_Q = \sum \limits_{ij} \frac{\partial Q}{\partial \alpha_i}
  (C_{\alpha}^{-1})_{ij} \frac{\partial Q}{\partial \alpha_j}.
\end{equation}
When using these values to obtain errors on the signal yield in any 
bin, we must consider the fact that the nature of our procedure leads to 
highly-correlated results for each event and its $n_c$ nearest neighbors; 
thus, simply
adding the $\sigma_{Q}$ values in quadrature would certainly underestimate the
true error. The actual degree of correlation of the $Q$-factors would depend on
the population of the bins. A safe choice is to assume 100\%
correlation; thus,
\begin{equation}
  \label{eq:q-err}
  \sigma_{\mathcal{Y}} = \sum\limits_{i}^{n_{bin}} \sigma_{Qi},
\end{equation}
which provides an overestimate of the true uncertainty inherent in the 
procedure.

In addition to the uncertainties associated with the fits, there will also be
a purely statistical error associated with the signal yield in each bin, given
by Poisson statistics. 
For large values of $\mathcal{Y}$, the signal yield obtained using 
(\ref{eq:sig-yield}), this can be taken to be $\sqrt{\mathcal{Y}}$; 
however, for smaller $\mathcal{Y}$ the upper limit is better approximated by
${1 + \sqrt{\mathcal{Y} + 0.75}}$~\cite{cite:poisson-errors}.
The total uncertainty on the signal yield in any bin is then obtained by adding
the fit errors, calculated using (\ref{eq:q-err}), in quadrature with the
statistical errors discussed above.
\begin{figure*}
  \begin{center}
  \includegraphics[width=1.0\textwidth]{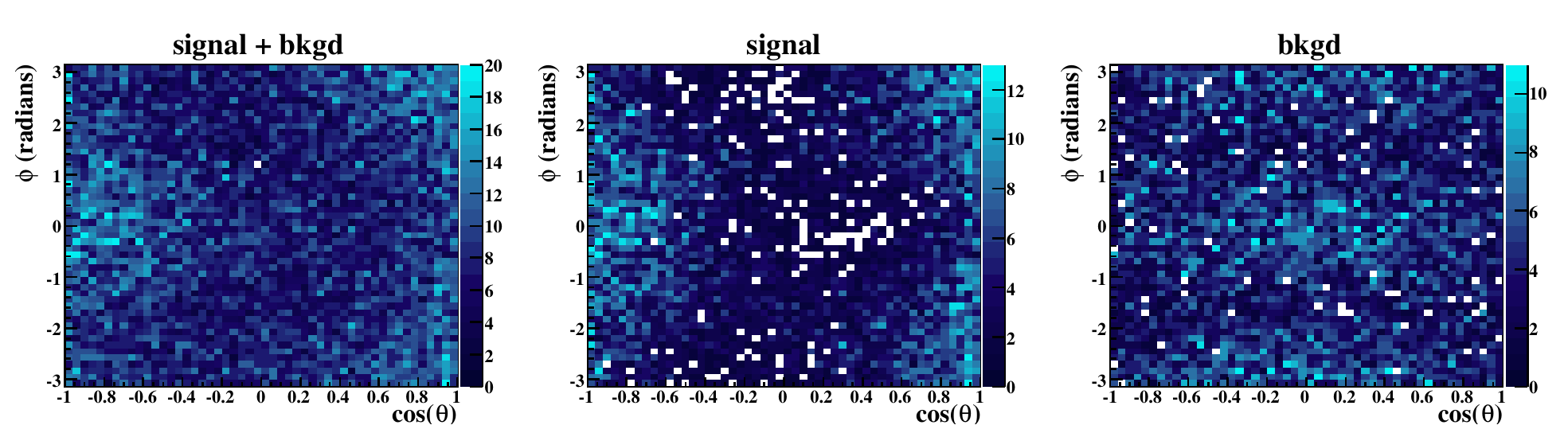}
  \caption[]{\label{fig:decay-angles-gen}
    (Color Online)
    $\phi$ (radians) vs $\cos{\theta}$: Generated decay angular distributions
    for all events (left), only signal events (middle) and only background
    events (right).
  }
  \end{center}
\end{figure*}

\section{\label{section:example}Example Application}

As an example, we will consider the reaction ${\gamma p \rightarrow p \omega}$
in a single $(s,t)$ bin, {\em i.e.} a single center-of-mass energy and 
production angle bin (extending the example to avoid binning in
production angle, or $t$, is discussed below). The $\omega$ decays to
$\pi^+\pi^-\pi^0$ about 90\% of the time; thus, we will assume we have a 
detector which has reconstructed ${\gamma p \rightarrow p \pi^+\pi^-\pi^0}$
events. Of course, there are production mechanisms other than  
${\gamma p \rightarrow p \omega}$ which can produce this final state and there
is no cut which can be performed to separate out events that originated from
${\gamma p \rightarrow p \omega}$. Below we will construct a toy-model of this
situation by generating Monte Carlo events for both signal, {\em i.e.}
$\omega$ events, and background, {\em i.e.} non-$\omega$ $\pi^+\pi^-\pi^0$
events (10,000 events were generated for each).
The goal of our model analysis is to extract the $\omega$ polarization
observables known as the {\em spin density matrix elements}, denoted by
$\rho^0_{MM'}$ (discussed below).

\begin{figure}
  \centering
  \includegraphics[width=0.45\textwidth]{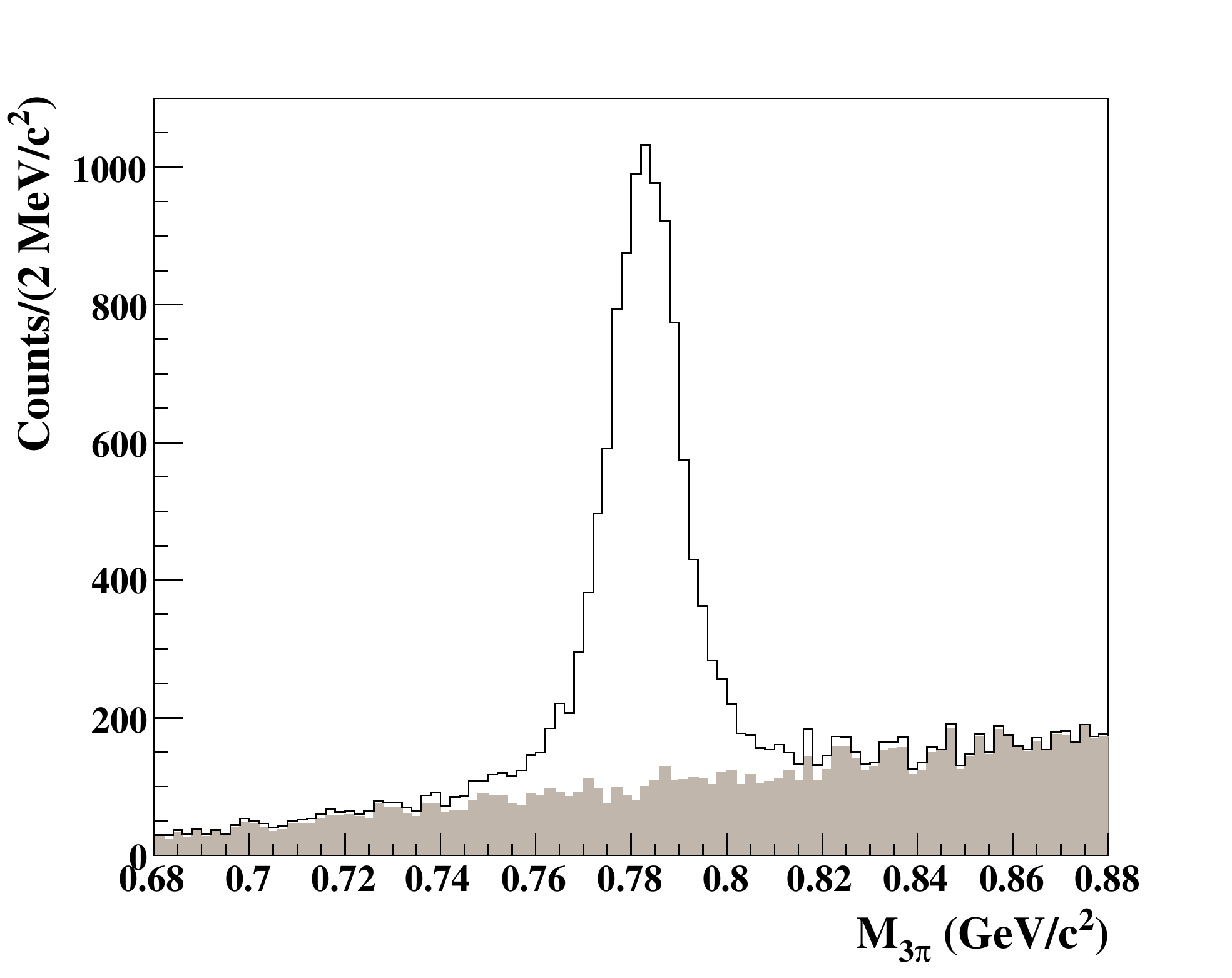}
  \caption[]{\label{fig:m3pi-gen}
    Mass of the $\pi^+\pi^-\pi^0$ system (GeV/c$^2$) for all generated events
    (unshaded) and for only the background (shaded).
  }
\end{figure}

In terms of the mass of the $\pi^+\pi^-\pi^0$ system, $m_{3\pi}$, the $\omega$
events were generated according to 3-body phase space weighted by a 
Voigtian (a convolution of a Breit-Wigner and a Gaussian, see (\ref{eq:voigt}))
to account for both the natural width of the $\omega$ and 
detector resolution. For this example, we chose to use $\sigma = 5$~MeV/c$^2$
for the detector resolution (see Figure~\ref{fig:m3pi-gen}). 
The goal of our analysis is to extract the three measurable elements of the
spin density matrix (for the case where neither the beam nor target are 
polarized) traditionally chosen to be $\rho^0_{00}$, $\rho^0_{1-1}$ and 
$Re\rho^0_{10}$. These can be accessed by examining the 
distribution of the decay products ($\pi^+\pi^-\pi^0$) of the $\omega$ in its
rest frame.

For this example, we chose to work in the helicity system which defines the
$z$ axis as the direction of the $\omega$ in the overall center-of-mass 
frame, the $y$ axis as the normal to the production plane and the $x$ axis is
simply given by ${\hat{x} = \hat{y} \times \hat{z}}$.
The decay angles $\theta,\phi$ are the polar and azimuthal angles of the normal
to the decay plane in the $\omega$ rest frame, {\em i.e.} the angles of the
vector ${\left(\vec{p}_{\pi^+} \times \vec{p}_{\pi^-}\right)}$.
The decay angular distribution of the $\omega$ in its rest frame is then 
given by~\cite{cite:schilling-1970}
\begin{eqnarray}
  \label{eq:schil}
  W(\theta,\phi) 
  = \frac{3}{4\pi} \left(\frac{1}{2}(1 - \rho^0_{00})
  + \frac{1}{2}(3\rho^0_{00} - 1)\cos^2{\theta}\right. 
  \nonumber\\
  - \left.\rho^0_{1-1}\sin^2{\theta}\cos{2\phi}
  - \sqrt{2}Re\rho^0_{10}\sin{2\theta}\cos{\phi}\right),
\end{eqnarray}
which follows directly from the fact that the $\omega$ is a vector particle;
it has spin-parity $J^P = 1^-$. 

We chose to use the following $\rho^0_{MM'}$ values for this example:
\begin{subequations}
  \label{eq:rho-gen}
  \begin{equation}
  \rho^0_{00} = 0.65 
  \end{equation}
  \begin{equation}  
  \rho^0_{1-1} = 0.05 
  \end{equation}
  \begin{equation}
    Re\rho^0_{10} = 0.10
  \end{equation}
\end{subequations}
The resulting generated decay distribution is shown in 
Figure~\ref{fig:decay-angles-gen}.

For the background, we chose to generate it according to 3-body phase space 
weighted by a linear function in $m_{3\pi}$ and 
\begin{equation}
  W(\theta,\phi) = \frac{1}{6\pi}\left(1 + |\sin{\theta}\cos{\phi}|\right)
\end{equation}
in the decay angles. Figure~\ref{fig:m3pi-gen} shows the $\pi^+\pi^-\pi^0$ mass
spectrum for all generated events and for just the background. The generated
decay angular distributions for all events, along with only the signal and
background are shown in Figure~\ref{fig:decay-angles-gen}. There is clearly no
way to separate out the signal events through the use of a cut.

\begin{figure*}
  \begin{center}
  \includegraphics[width=1.0\textwidth]{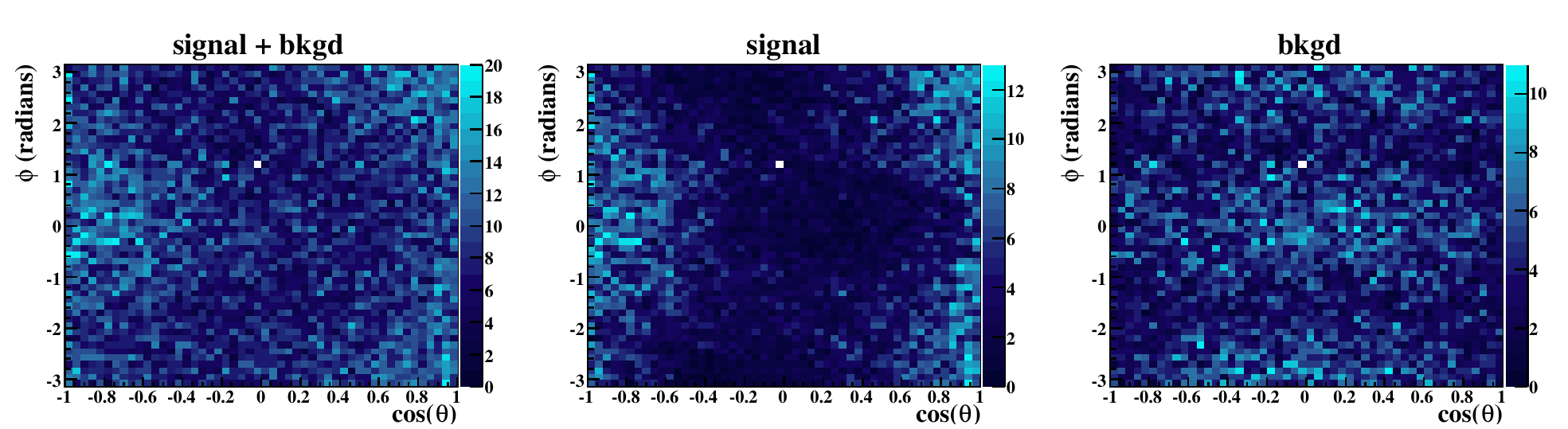}
  \caption[]{\label{fig:decay-angles-fit}
    (Color Online)
    $\phi$ (radians) vs $\cos{\theta}$: Extracted decay angular distributions
    for all events (left), for events weighted by $Q$, signal (middle), and
    for events weighted by $1-Q$, background (right).
  }
  \end{center}
\end{figure*}

\subsection{Applying the Procedure}
\label{section:example:apply}

To obtain the $Q$-factors, we first need to identify the relevant coordinates,
{\em i.e.} the kinematic variables in which we need to separate signal 
from background. The $\pi^+\pi^-\pi^0$ mass will be used as the reference 
coordinate, $\xi_r \equiv m_{3\pi}$.
The stated goal of our analysis is to extract the 
$\rho^0_{MM'}$ elements. We will do this using (\ref{eq:schil}); thus, only
the angles $\theta,\phi$ are relevant. Other decay variables, such as the
distance from the edge of the $\pi^+\pi^-\pi^0$ Dalitz plot, are not relevant
to this analysis --- though, they would be in other analyses 
(see Section~\ref{section:example:full-pwa}). 

Using the notation of
Section~\ref{section:method}, $\vec{\xi} = (m_{3\pi},\cos{\theta},\phi)$ and
the distance between any two points, $d_{ij}$, is given by
\begin{equation}
  d^2_{ij} = \left(\frac{\cos{\theta_i} - \cos{\theta_j}}{2}\right)^2 
  + \left(\frac{\phi_i - \phi_j}{2\pi}\right)^2.
\end{equation}
The functional dependence of the signal and background on the reference 
coordinate, $m_{3\pi}$, are
\begin{subequations}
\begin{equation}
  F_s(m_{3\pi},\vec{\alpha}) = 
  s\cdot V(m_{3\pi},m_{\omega},\Gamma_{\omega},\sigma)
\end{equation}
\begin{equation}
  F_b(m_{3\pi},\vec{\alpha}) = b_1 m_{3\pi} + b_0,  
\end{equation}
\end{subequations}
where $m_{\omega} = 0.78256$~GeV/c$^2$, $\Gamma_{\omega} = 8.44$~MeV, 
${\sigma=5}$~MeV is the simulated detector resolution, 
${\vec{\alpha} = (s,b_1,b_0)}$ are unknown parameters and 
\begin{eqnarray}
  \label{eq:voigt}
  V(m_{3\pi},m_{\omega},\Gamma_{\omega},\sigma)  
  = \hspace{0.225\textwidth}\nonumber \\
  \frac{1}{\sqrt{2\pi}\sigma} Re \left[
    w\left(\frac{1}{2\sqrt{\sigma}}(m_{3\pi} - m_{\omega}) 
    + i\frac{\Gamma_{\omega}}{2\sqrt{2}\sigma}
    \right)\right],
\end{eqnarray}
is the convolution of a Gaussian and non-relativistic Breit-Wigner
known as a Voigtian ($w(z)$ is the complex error function). 

As stated above, the number of nearest neighbor events required depends on the
analysis. Specifically, it depends on how many unknown parameters there are, 
along with the functional forms of $F_s$ and $F_b$. For this relatively simple
case, the value $n_c = 100$ works well. For each simulated event, we then
find the $n_c$ closest events (containing both signal and background) and 
perform an unbinned maximum likelihood fit, using the CERNLIB package 
MINUIT~\cite{cite:minuit}, to determine the estimators $\hat{\alpha}$.  
The $Q$-factors are then calculated from (\ref{eq:q-factor})
and the uncertainties are straightforward to calculate using 
Section~\ref{section:errors}.

\begin{figure}
  \centering
  \includegraphics[width=0.45\textwidth]{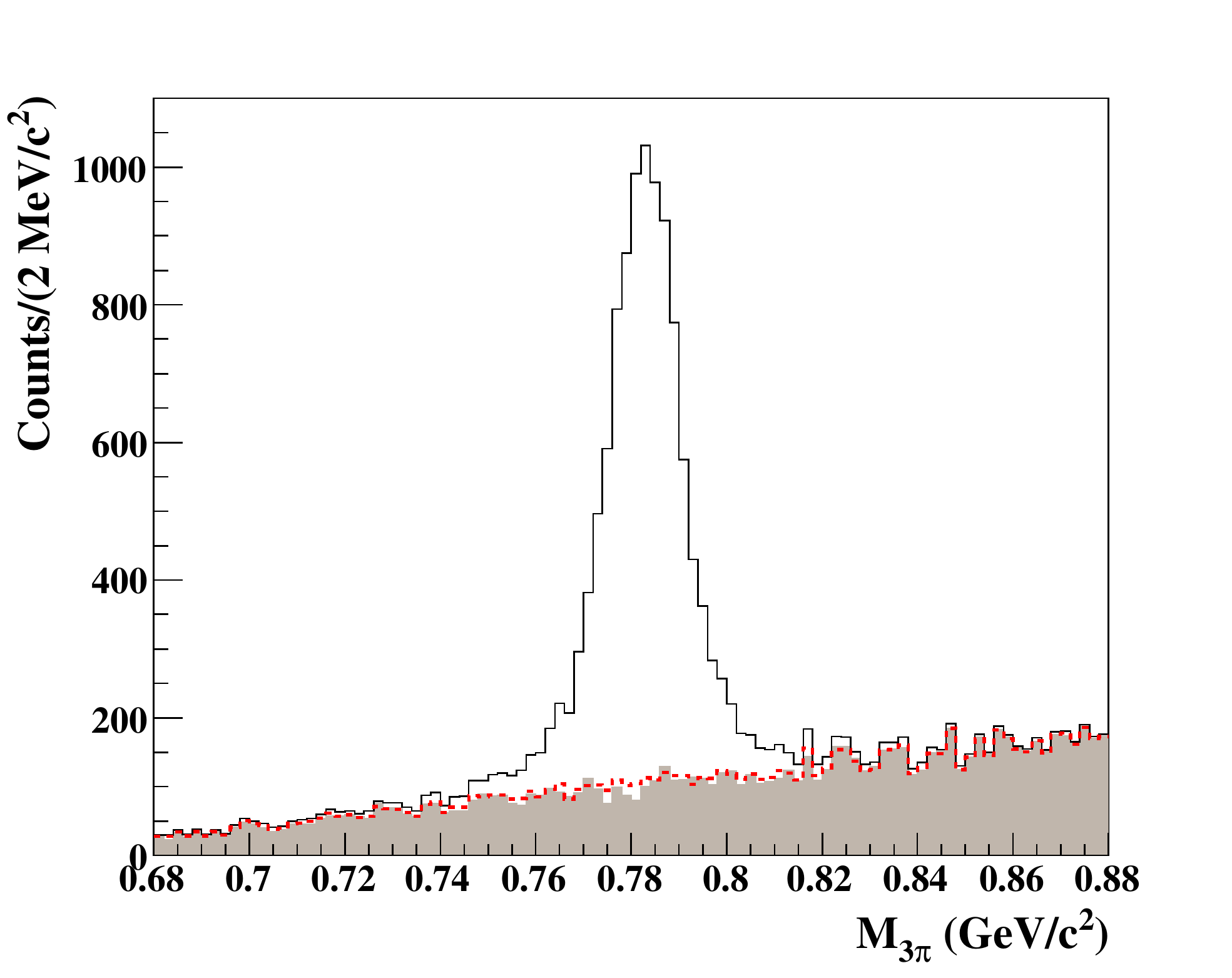}
  \caption[]{\label{fig:m3pi-fit}
    (Color Online)
    Mass of the $\pi^+\pi^-\pi^0$ system (GeV/c$^2$) for all generated events
    (unshaded), only generated background events (shaded) and all generated
    events weighted by $1-Q$ (dashed-red).
  }
\end{figure}

\begin{figure*}
  \centering
  \subfigure[]{
    \includegraphics[width=0.4\textwidth]{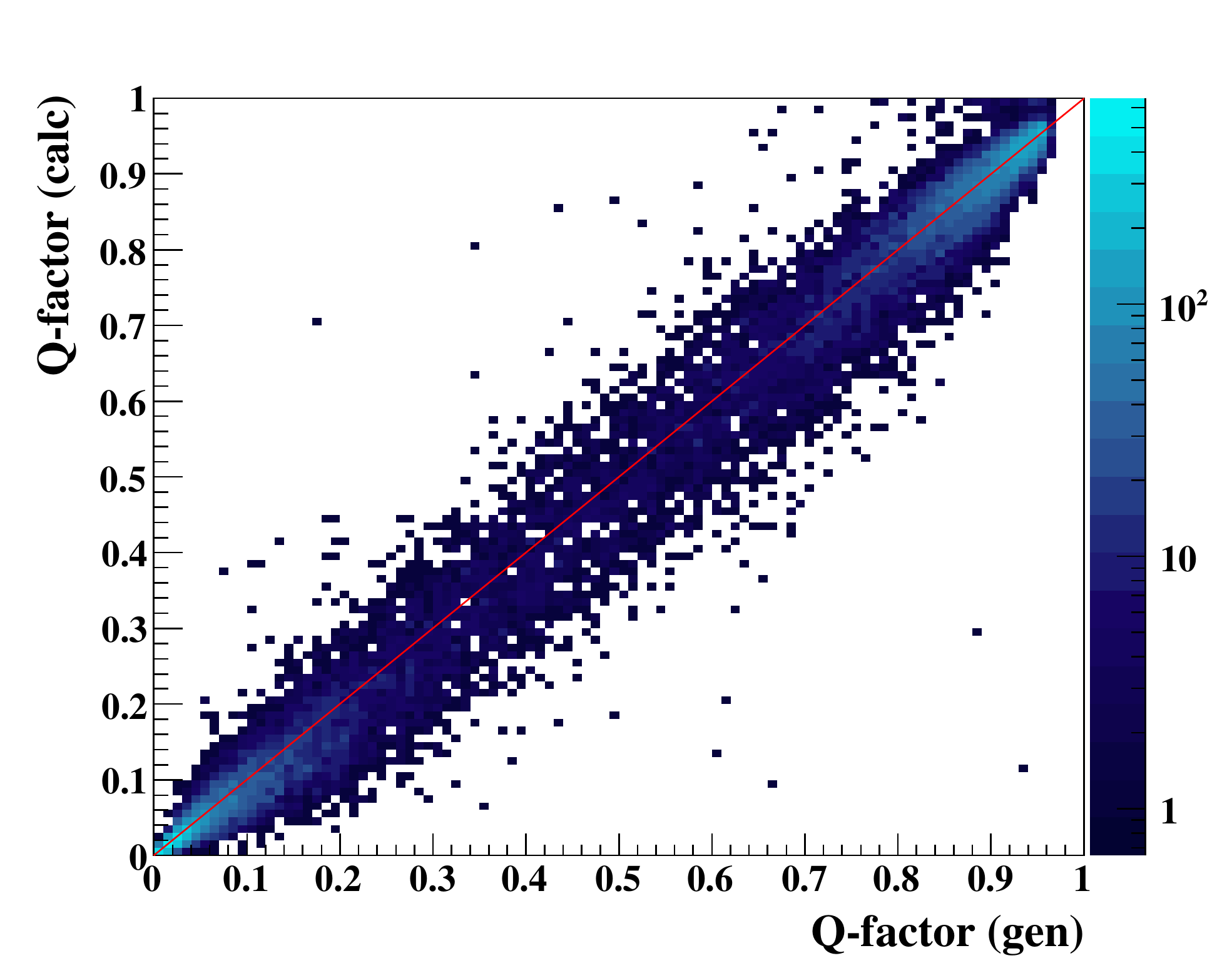}
  }
  \hspace{0.1\textwidth}
  \subfigure[]{
    \includegraphics[width=0.4\textwidth]{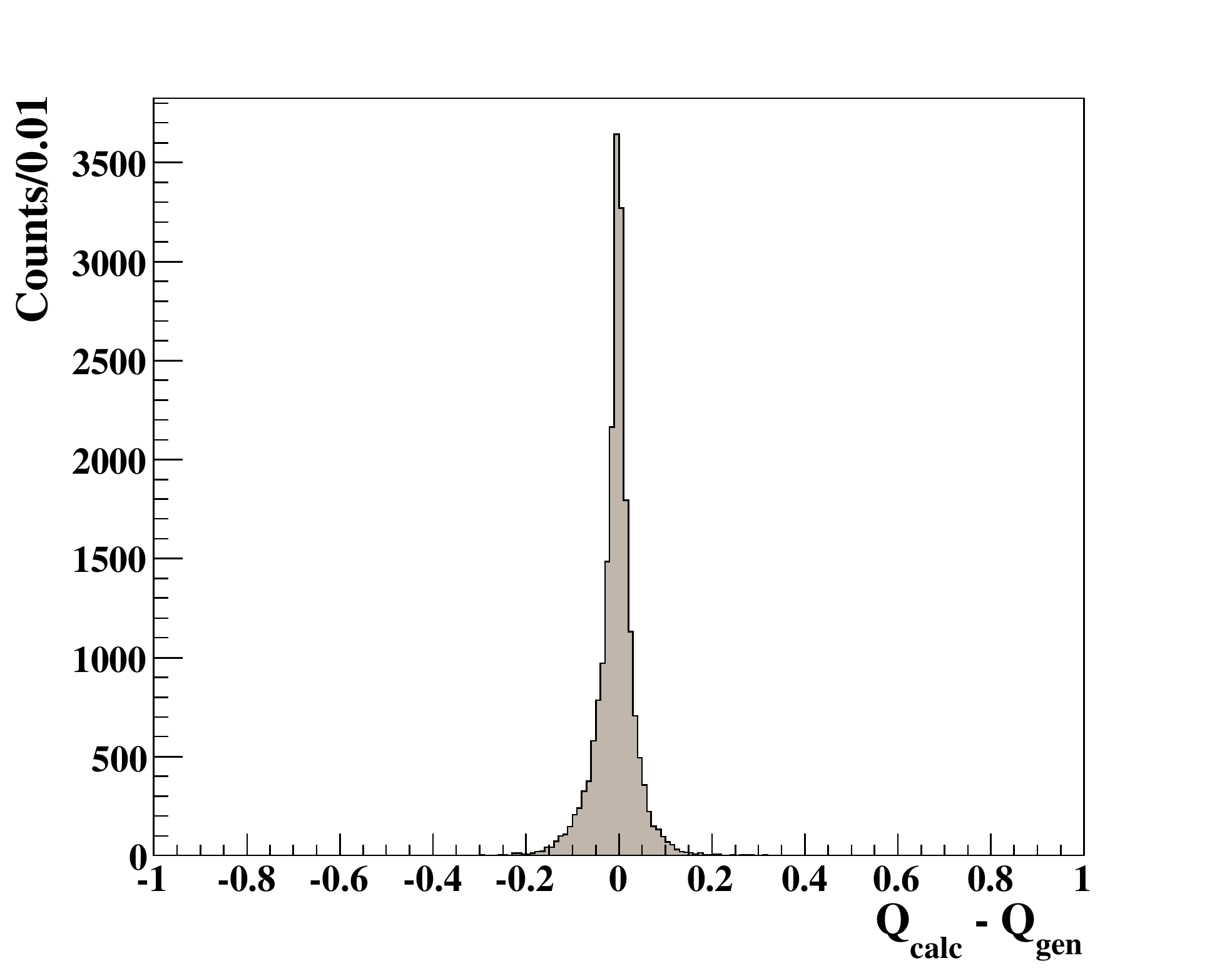}
  }
  \caption[]{\label{fig:q-compare}
    (Color Online)
    (a) Calculated $Q$-factors vs generated $Q$-factors. The red line indicates
    where ${Q_{calc} = Q_{gen}}$.
    (b)~${Q_{calc} - Q_{gen}}$: The difference between the generated and 
    calculated $Q$-factors.
  }
\end{figure*}

Figure~\ref{fig:m3pi-fit} shows the comparison of the extracted and generated 
background $m_{3\pi}$ distributions integrated over all decay angles. The
agreement is quite good; however, we are looking for more than just global 
agreement. Figure~\ref{fig:decay-angles-fit} shows the extracted angular 
distributions for the signal and background. The agreement with the 
generated distributions is excellent (see Figure~\ref{fig:decay-angles-gen}) .
We can also compare the $Q$-factors extracted by the fits to the theoretical 
distributions from which our data was generated. 
Figure~\ref{fig:q-compare} shows that the extracted values are in very good
agreement with the generated values.

We conclude this section by discussing the importance of quality control in
the fits. For this example, we performed 20,000 independent fits to extract the
$Q$-factors. To avoid problems which can arise due to fits 
not converging or finding local minima, each unbinned maximum likelihood fit
was run with three different sets of starting values for the parameters
$\vec{\alpha}$: 
(1)~100\% signal; 
(2)~100\% background; 
(3)~50\% signal, 50\% background.
In all cases, the fit with the best likelihood was used.
The $n_c$ events were then binned and a $\chi^2/ndf$ was obtained. 
In about 2\% of the fits the $\chi^2/ndf$ was very large, 
a clear indicator that the
fit had not found the best estimators $\hat{\alpha}$.
For these events, a binned $\chi^2$ fit was run to obtain the $Q$-factor. 

\subsection{Examining the Errors}

As discussed in Section~\ref{section:errors}, the covariance matrix obtained
from each event's fit can be used to obtain the uncertainty in $Q$, 
$\sigma_Q$, using (\ref{eq:sigma-q}). The nature of our procedure leads to a
high degree of correlation between neighboring event's $Q$-factors. This means
that adding the uncertainties in quadrature would definitely 
underestimate the true error.
In Section~\ref{section:errors}, we argued that a better approach was to 
assume 100\% correlation which provides an overestimate of the true error.

\begin{figure*}
  \centering
  \subfigure[]{
    \includegraphics[width=0.4\textwidth]{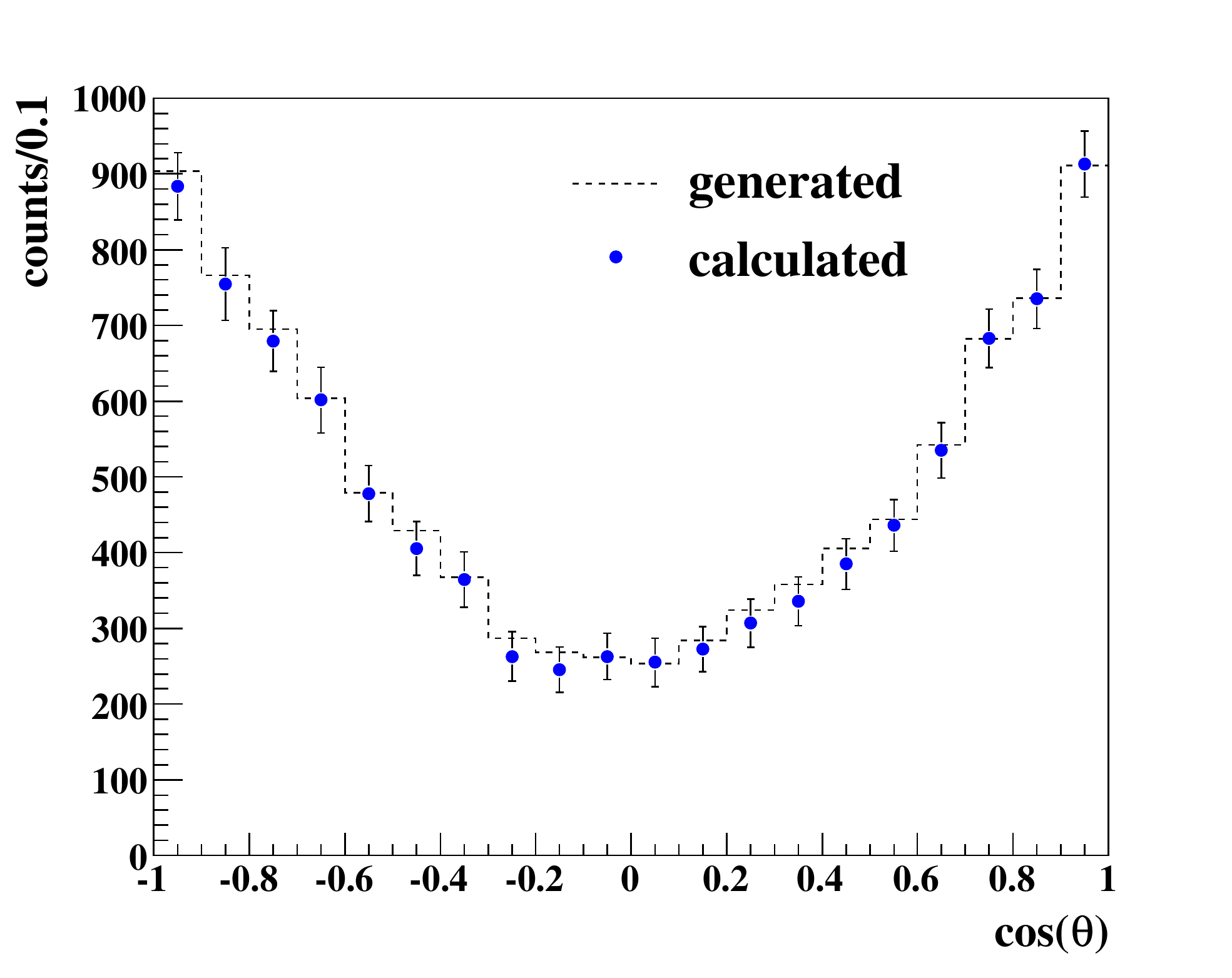}
  }
  \hspace{0.1\textwidth}
  \subfigure[]{
    \includegraphics[width=0.4\textwidth]{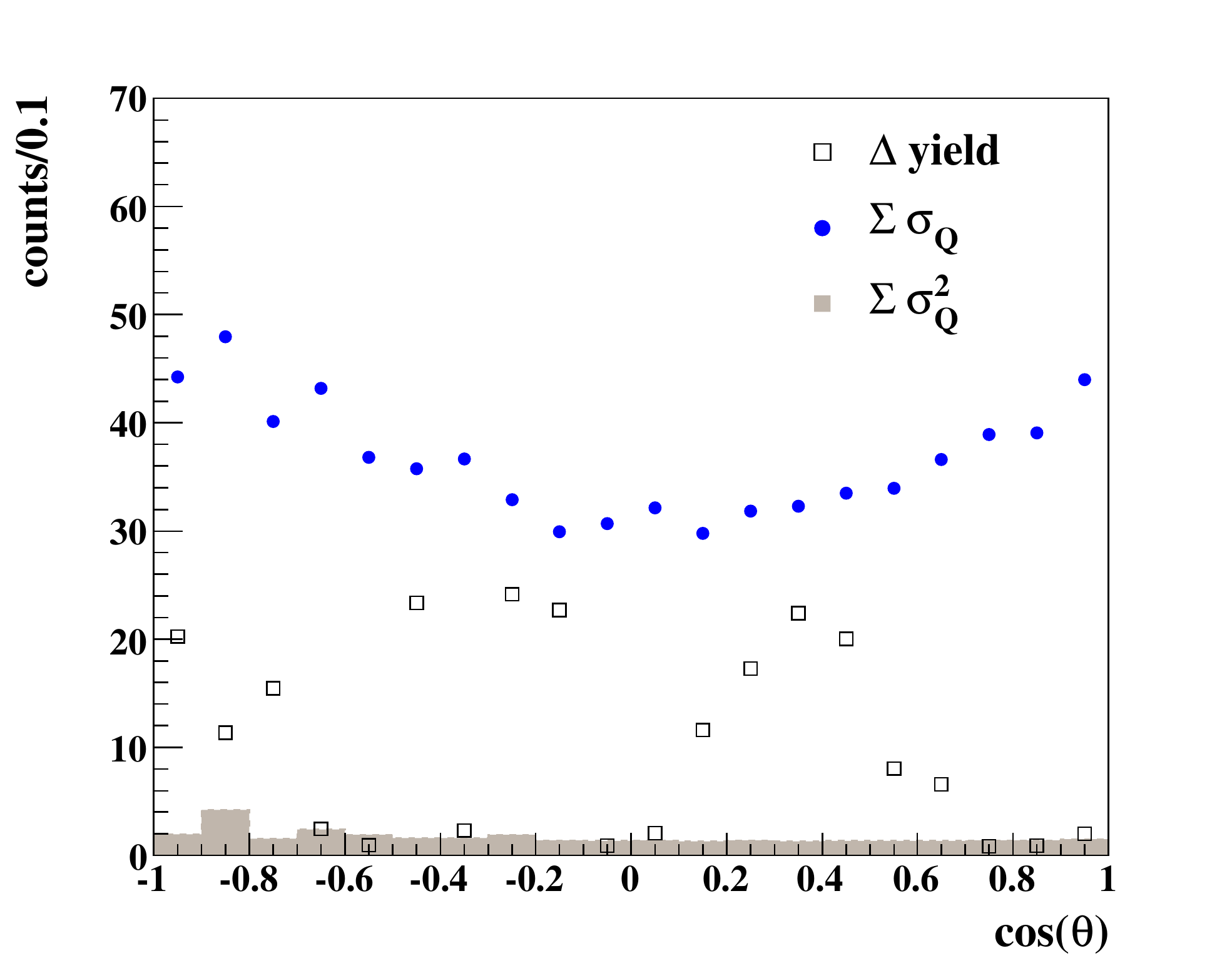}
  }
  \caption[]{\label{fig:compare-q-err}
    (Color Online)
    (a) Signal yield vs $\cos{\theta}$ for generated signal events (dashed) and
    all generated events weighted by $Q$-factors (blue circles). The error bars
    on the extracted yields were obtained using (\ref{eq:q-err}).
    (b) Comparison of the true error on the signal yield, 
    $\Delta \mathcal{Y} = |\mathcal{Y}_{gen} - \mathcal{Y}_{calc}|$
    (open black squares), to the
    error bars obtained using the $\sigma_Q$ values obtained from 
    (\ref{eq:sigma-q}) combined assuming 100\% correlation (blue circles) and
    assuming no correlation (shaded band at bottom).
  }
\end{figure*}

To examine the error bars in our toy example, we chose to project our data into
a one-dimensional distribution in $\cos{\theta}$. This was done to avoid 
bin occupancy issues which arise in the two-dimensional case due to limitations
in statistics.
Figure~\ref{fig:compare-q-err}(a) shows the comparison between the generated
and calculated $\cos{\theta}$ distributions. The agreement is excellent.
The error bars on the calculated
points were obtained using (\ref{eq:q-err}). For this study, we ignore 
the Poisson statistical uncertainty in the yield due to the fact that the 
number of generated events is known. In a real world analysis, these should be
included in the quoted error bars.

We can examine the quality of the error estimation by examining the difference
between the generated and calculated yields in each bin, $\Delta \mathcal{Y}$.
Figure~\ref{fig:compare-q-err}(b) shows the comparison between 
$\Delta \mathcal{Y}$, $\sigma_{\mathcal{Y}}$ obtained assuming 100\% 
correlation and $\sigma_{\mathcal{Y}}$ obtained assuming no correlation,
{\em i.e.} adding the individual uncertainties in quadrature.
As expected, the correlated errors provide an overestimate of 
$\Delta \mathcal{Y}$ in every bin, while 
the uncorrelated errors greatly underestimate $\Delta \mathcal{Y}$ in the 
majority of bins.

It is not surprising that the errors obtained using (\ref{eq:q-err}) are 
better estimates of the true uncertainty when the bin occupancy is closer to
$n_c$. We could come up with a scheme to scale the size of the error bars 
according to bin occupancy; however, recall that in this example we fit 
the background with the functional form to which it was generated. This 
typically will not be the case for real data. In a real analysis, fitting
with different functional
forms and examining the extracted signal 
distributions to estimate systematic effects due to the choice of background 
is generally required. In this case, one should simply consider that the 
uncorrelated error bars are overestimates for each choice of background 
function when estimating the systematic uncertainty in the signal yields.


\subsection{Extracting Observables: Event-Based Fitting using $Q$-Factors}
\label{section:example:rho}

The goal of our model analysis is to extract the spin density matrix elements.
When studying multi-dimensional problems, binning the data is often 
undesirable due to limitations in statistics. 
The $Q$-factors obtained for each event above can be used in conjunction with
the unbinned maximum likelihood method to avoid this difficulty.

If we could cleanly separate out the signal events from the background, then 
the likelihood function would be defined as
\begin{equation}
  \mathcal{L} = \prod\limits_i^{n_s} W(\theta_i,\phi_i),
\end{equation}
where $W$ is the decay angular distribution defined in (\ref{eq:schil}).
We could then obtain estimators for the spin density matrix elements by 
minimizing
\begin{equation}
  \label{eq:log-L}
  -\ln{\mathcal{L}} = -\sum\limits_i^{n_s} \ln{W(\theta_i,\phi_i)}.
\end{equation}
In this example, it is not possible to separate the signal and background 
event samples; however, we can use the $Q$-factors to achieve the same effect
by rewriting (\ref{eq:log-L}) as
\begin{equation}
  \label{eq:log-L-Q}
  -\ln{\mathcal{L}} = -\sum\limits_i^{n} Q_i \ln{W(\theta_i,\phi_i)},
\end{equation}
where the sum is now over all events (signal and background).
Thus, the $Q$-factors are used to weight each event's contribution to the 
likelihood.

Using the $Q$-factors obtained in Section~\ref{section:example:apply}, 
minimizing (\ref{eq:log-L-Q}) yields
\begin{subequations}
  \begin{equation}
  \rho^0_{00} = 0.654\pm0.011 
  \end{equation}
  \begin{equation}
  \rho^0_{1-1} = 0.046\pm0.008 
  \end{equation}
  \begin{equation}
  Re\rho^0_{10} = 0.099\pm0.007,
  \end{equation}
\end{subequations}
where the uncertainties are purely statistical (obtained from the fit 
covariance matrix). Thus, the values extracted for the spin density matrix
elements are in excellent agreement with the values used to generate the
data given in (\ref{eq:rho-gen}).

\subsection{Extending the Example}
\label{section:example:full-pwa}

To extend this example to allow for the case where the data is not binned in
production angle, we would simply need to include $\cos{\theta^{\omega}_{CM}}$
or $t$ in the vector of relevant coordinates, $\vec{\xi}$. 
To perform a full partial wave analysis on the data, we would also need to 
include any additional kinematic variables which factor into the partial wave 
amplitudes, {\em e.g.} the distance from the edge of the $\pi^+\pi^-\pi^0$ 
Dalitz plot (typically included in the $\omega$ decay amplitude). We would then
construct the likelihood from the partial waves and minimize 
$-\ln{\mathcal{L}}$ using the $Q$-factors obtained by applying our procedure 
including the additional coordinates. An example of this can be found 
in~\cite{cite:williams-thesis}.
\section{\label{section:conclusions}Conclusions}

In this paper, we have presented a procedure for separating signals from 
non-interfering backgrounds by determining, on an event-by-event basis, a 
quality factor ($Q$-factor) that a given event originated from the signal
distribution. We have shown that this $Q$-factor can be used as an event 
weight in subsequent analysis procedures to allow more direct access to the
true signal distribution. Though this procedure may be computationally 
expensive, in principle it only needs to be performed once for each event in
the data sample.
\begin{center}
\mbox{}\\\vspace{0.2in}
{\normalsize \textbf{Acknowledgments}}
\end{center}
\vspace{0.2in}
This work was supported by grants from the United States Department of Energy
No. DE-FG02-87ER40315 and 
the National Science Foundation No. 0653316 through the 
``Physics at the Information Frontier'' program.



\end{document}